# Doppler-free selective reflection spectroscopy of electric-quadrupole transitions

*Eng Aik Chan*[1], *Syed Abdullah Aljunid*[1], Athanasios Laliotis[2], David Wilkowski[1,3,4], Martial Ducloy[2,*]

[1]Centre for Disruptive Photonic Technologies, The Photonics Institute, Nanyang Technological University, Singapore 637371, Singapore.
[2]Laboratoire de Physique des Lasers, Université Sorbonne Paris Nord, F-93430 Villetaneuse, France.
[3]Centre for Quantum Technologies, National University of Singapore (NUS), Singapore 117543, Singapore.
[4]MajuLab, International Joint Research Unit IRL 3654, CNRS, Université Côte d'Azur, Sorbonne Université, National University of Singapore, Nanyang Technological University, Singapore.

*martial.ducloy@univ-paris13.fr

**Electric-dipole-forbidden transitions play an important role as in quantum sensing, quantum information, and fundamental test in physics. As such, the development of novel and sensitive spectroscopic methods is of major interest. Here, we present a Doppler-free selective reflection experiment on the $6^2S_{1/2} \rightarrow 5^2D_{5/2}$ electric-quadrupole transition of cesium vapor at the vicinity of a sapphire window. This is achieved by a precision experiment overcoming limitations due to the small signal amplitude of forbidden transitions. Narrow sub-Doppler lines allow for a collisional broadening measurement on the electric-quadrupole line. The interaction of cesium atoms with the sapphire surface of the cell is evidenced, but, due to its weak contribution, a quantitative analysis remains challenging. Nevertheless, our experiment paves the way for further studies of the Casimir-Polder interaction between exotic excited-state atoms and dielectric surfaces.**

With the progress of laser sources, spectroscopic methods have undergone a fast development, in particular the devising of Doppler-free spectroscopic methods for the study of atomic and molecular gases (saturated absorption, pump-probe spectroscopy, two-photon absorption…) [1], [2]. Doppler-free approaches give access to fundamental properties of atoms and molecules such as high-resolution spectra, precise measurements of emission wavelengths and level structures including Rydberg states, collisional line broadenings and shifts. Most of the spectroscopic studies have been performed using electric-dipole transitions. But, electric-dipole-forbidden lines also find interest, in particular in tabletop physics experiments for time keeping [3], probing angular momentum of photons [4], [5], [6] thus testing the fundamental symmetries of nature [7] [8] [9].

Many of the above applications require Doppler-free precision spectroscopy. However, saturated absorption techniques have proved elusive in the cases of forbidden transitions due to high saturation intensities. For this reason alternative techniques have been developed particularly on electric-quadrupole (E2) transitions [10], [11], [12]. Those recent demonstrations involve a three-level system consisting of a strong electric-dipole (E1) transition and a weak E2 transitions, allowing an investigation of optical pumping mechanisms in E2 transitions [12].

Another Doppler-free spectroscopy approach is in linear Selective Reflection (SR) spectroscopy which does not require saturation mechanisms, and thus high optical powers, to achieve the Doppler-free regime. In SR, one monitors the changes in the optical reflectivity at a dielectric-window/atomic-vapor interface due to the modification of the real part of the vapor refractive index when the light frequency crosses an atomic resonance. Because of collisions with the surface, the atomic response is in the transient regime resulting to sub-Doppler features [13], [14], [15] in the SR spectrum. Due to the logarithmic singularity nature of the sub-Doppler peak, SR at normal incidence becomes fully Doppler-free after deploying a frequency modulation (FM) technique [15], [16], [17]. Up to now, Frequency Modulated Selective Reflection (FMSR) has been mainly applied on atomic electric-dipole-allowed transitions, to analyze long- and short-range atom-surface interactions with flat surfaces (attraction and repulsion, influence of surface excitations) [18], [19], [20] or metamaterials [21], collisional line broadenings and shifts [22], [23], surface-induced symmetry breaking and energy level mixing, etc. (see *e.g.* [24], [25] and ref. therein). More recently the FMSR technique has also found applications in rovibrational molecular spectroscopy [26].

In this Letter we present an extension of FMSR spectroscopy to E2 lines [27], [28], [29]. In the same way that vapor absorption has been observed around E2 lines, the contribution of E2 transitions to the vapor index implies reflection changes at dielectric-vapor interfaces. We show that it gives access to pressure broadenings and allows us to explore the influence of surface-induced potential, paving the way for QED tests with forbidden transitions. Contrary to nonlinear spectroscopic schemes, linear selective reflection allows sub-Doppler spectroscopy with a one beam set-up using non-saturating optical power. The very small signals, due to the weak transition strength of E2 transitions, are obtained using a stable set-up allowing for integration times of typically a few days.

To investigate the SR spectroscopy on an E2 transition, we address the $6^2S_{1/2} - 5^2D_{5/2}$ transition of cesium at 685nm which has an electric-quadrupole matrix element of $\langle e||Q||g\rangle \sim 42$ a.u., leading to an Einstein coefficient of $A_{eg} \sim 2\pi \times 3.5$ Hz [30], [31]. The excited state lifetime is about $\tau = 1280$ ns, dominated by the electric-dipole $5^2D_{5/2}$ -> $6^2P_{3/2}$ transition, giving an averaged saturation intensity of about $2W/cm^2$ [12]. The $5^2D_{5/2}$ hyperfine structure, and the experimentally relevant relative transition probabilities from the state $|6^2S_{1/2}, F=4\rangle$ given by

$$S^Q_{FF'} = (2F'+1)(2J+1)\begin{Bmatrix} J & J' & 2 \\ F' & F & I \end{Bmatrix}^2 \qquad (1)$$

are shown on Fig. 1. Here I = 7/2 is the nuclear spin.

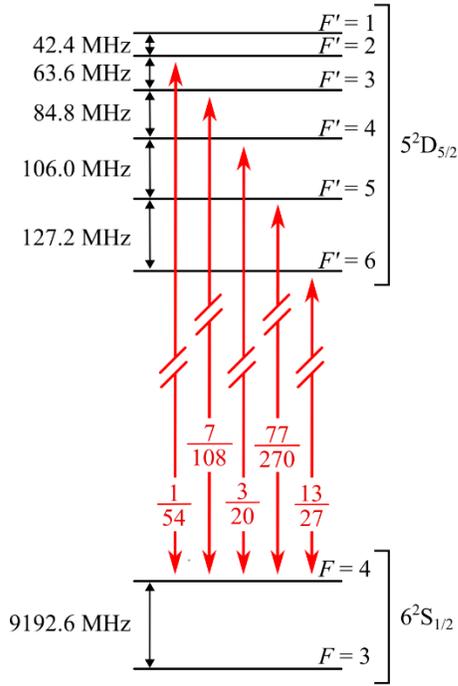

Figure 1. The 685 nm $6^2S_{1/2}$-$5^2D_{5/2}$ hyperfine transition (red-vertical arrows) and the relative hyperfine transition probabilities, $S_{FF'}$ (Eq. 1) of the allowed transitions ($\Delta m = \pm 2, \pm 1, 0$), from the $F = 4$ ground hyperfine level.

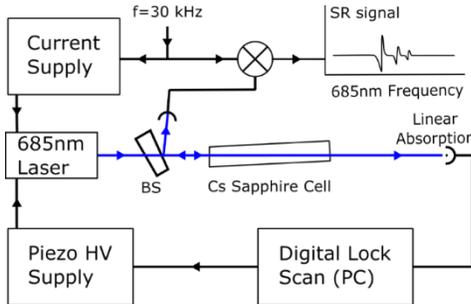

**Figure 2.** Schematic of the experimental setup. The current to the laser diode is modulated at 30kHz to generate the frequency sidebands used for FMSR spectroscopy. The cesium sapphire cell is heated by an oven up to a temperature of 285°C. The transmitted beam is used as a frequency reference and for cesium density calibration by direct absorption spectroscopy. The reflected beam signal of the sapphire-cesium vapor interface is demodulated before being stored on a PC and averaged over few days of acquisitions. BS: beam splitter.

A schematic of the experimental setup is depicted in Fig. 2. The 685 nm light source is a single-frequency external cavity diode laser (ECDL) with an output power of 11 mW. The current of the diode laser is modulated to provide a frequency-modulation-amplitude depth of 10 MHz at a 30 kHz rate. A non-saturating laser beam, with intensity on the order of $1W/cm^2$, is sent onto an 8 cm-long spectroscopic cesium cell, with sapphire windows on both sides. The cell is placed inside an oven with an adjustable temperature up to T = 285°C. The cell temperature is almost homogeneous with a slightly higher temperature at the input and output windows to avoid unwanted cesium condensation on the laser path.

The ECDL cavity length and the diode laser current are slowly scanned (period ~ 1 s) over a frequency range of 3 GHz that covers the Doppler-broadened spectrum of the F = 4 -> F' = 2, 3, 4, 5, 6 transitions of the E2-line. The transmitted signal through the cesium cell measures the Doppler-broadened absorption at 685nm and provide a frequency reference for the SR spectra, as discussed below. In addition, the gas temperature and the atomic density are derived from the transmission spectra. The latter are obtained using a fit with Gaussian function for each allowed hyperfine transition with a relative weight given by eq. (1) and frequency separations given in Fig. 1. From the same transmission data, we extract the atomic density and the gas pressure, following [12]. In Fig. 3 (a) and (b), we show two spectra (red curves) corresponding to temperatures of 223°C and 260°C respectively, along with the corresponding fits (dashed-blue curves).

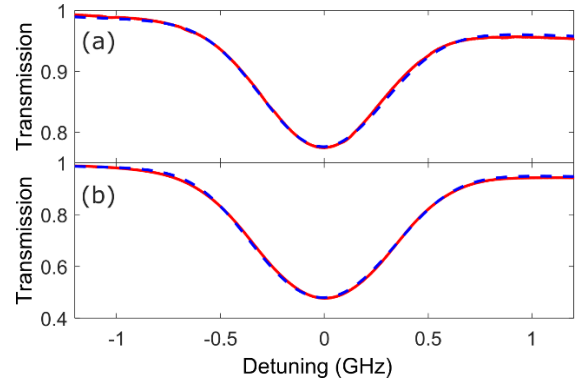

**Figure 3.** Transmission spectra data (red curve) and fit (dashed-blue curve) of the thermal gas for (a) $T$ = 223 °C (b) $T$ = 260 °C. The fit corresponds to a Gaussian function for each hyperfine transition with a relative weight and separation given in Fig. 1. In addition, a linear tilt is added to the fit to take into account the variation of the laser power when the frequency is scanned using the diode laser current and the cavity length.

The FMSR signal corresponding to light beam reflected from the internal window/vapor interface is collected on a detector followed by a lock-in amplifier referenced on the frequency of the diode current modulation.

The FMSR signal is proportional to the frequency derivative of the vapor effective susceptibility, $\bar{\chi}$, which is in turn proportional to the vapor density $N$ and the ratio between the Einstein coefficient $A_{eg}$ and the homogeneous transition linewidth $\gamma_c$ ($= 1/\tau$ at low density) [14], [32]. The FMSR signal, given by the frequency derivative of the relative variation of the reflectivity at the vapor/window interface $\Delta R(\omega)$, normalized to the off-resonant value of the reflectivity, R, is given by [32]:

$$I_{FMSR} = \frac{d(\Delta R/R)}{d\omega} = -\frac{2n}{(n^2-1)} Re\left(\frac{d\bar{\chi}}{d\omega}\right) \quad (2)$$

where $n$ is the refractive index of the window. The above equation is valid when the FM amplitude is smaller than the homogeneous transition linewidth. Furthermore, in the large Doppler-broadening approximation ($ku \gg \gamma_c$), the frequency derivative of the effective susceptibility is given by:

$$\frac{d\bar{\chi}}{d\omega} = -\frac{3A_{eg}k^3}{4\sqrt{\pi}} \frac{iN}{\varepsilon_o} \frac{1}{\gamma_c(ku)} I\left(\frac{2\omega}{\gamma_c}, A\right) \quad (3)$$

Here $u$ is the thermal velocity and $k = \frac{2\pi}{\lambda}$ is the laser field wavenumber, while the product $ku$ represents the thermal width of the transition. $I\left(\frac{2\omega}{\gamma_c}, A\right)$ is the SR lineshape that depends on a dimensionless parameter $A = \frac{2C_3 k^3}{\gamma_c}$. The SR lineshape can be found in [32]. $C_3$ is a coefficient defining the Casimir-Polder interaction energy between the atom and the dielectric window, whose distance dependence in the non-retarded approximation is given by $-\frac{C_3}{z^3}$ [25]. In the case of spectroscopic experiments, only the relative $C_3$ coefficient between the excited and ground states can be measured.

For this experiment, our cell is operated at high temperatures providing a cesium density in the range of $N = 4x10^{15} - 3x10^{16}$ cm$^{-3}$, corresponding to pressures in the one Torr range. In Fig. 4, we show SR spectra on the quadrupole $6S_{1/2}$—$5D_{5/2}$ transition taken at two different temperatures and densities. It is worth noting that the experimentally measured amplitude of the spectra (of the largest peak F = 4—F' = 6) is around $3\times10^{-8}$ (30 parts per billion) and roughly independent of the atomic density. This suggests that our experiment is in a regime dominated by pressure broadening and therefore an increase in density (accompanied by a linear increase in linewidth) has no significant visible effects on the amplitude [see eqs. (2) & (3)]. In Fig.4, we also show the corresponding fit of the FMSR spectra (dashed-blue curves), using eqs. (2) & (3) including Casimir-Polder interactions [32]. The free parameters of the fit are the ground-excited state relative $C_3$ coefficient, the collisional broadening $\gamma_c$ and shift, the signal amplitude, and a residual offset. Only the prominent F = 4 – F'= 6, 5, 4 hyperfine transitions are visible and fitted. Their relative amplitudes are left as free-fit parameters. The fitted relative amplitudes and the theory prediction are summarized in Table 1. At low temperature and pressure, the result agrees with the theory expectations, confirming the linearity of the SR spectroscopy. At larger pressure, some deviations are observed.

|  | F = 4—F' = 5 | F = 4—F' = 4 |
|---|---|---|
| Theory | 0.59 | 0.31 |
| Exp. at $T$ = 223°C | 0.56(10) | 0.36(10) |
| Exp. at $T$ = 260 °C | 0.38(10) | 0.43(10) |

**Table 1:** Relative amplitude ratio of the FMRS signal for the observed hyperfine transitions. The normalization is done with respect to the F = 4—F' = 6 transition.

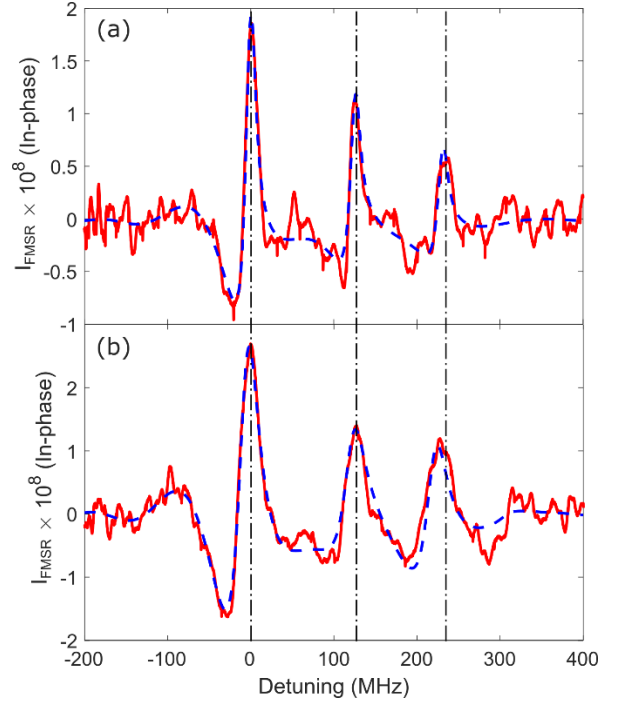

**Figure 4.** FMSR in-phase data (red curve) and fit (dashed-blue curve) for (a) $T$ = 223 °C (b) $T$ = 260 °C. The signal corresponds to the frequency derivative of the relative variation of the reflectivity $R$, obtained after demodulation. The FMSR in-quadrature data does not show measurable signal. Each experimental curve corresponds to 60 hours of aquisition. Only the peaks corresponding to F=4->F'=6,5,4 are visible and fitted with the expression given in eq. (2). The free parameters are; a global homogenous width $\gamma_c$ and C$_3$ coefficient, the amplitude of each hyperfine transition, a global detuning (frequency shift) in the ±15MHz range from the bulk spectroscopy centres. The vertical dashed-dotted black lines indicate the relative position of the hyperfine transition. The origin of the detuning corresponds to the F = 4 – F' = 6 hyperfine transition.

Collisional broadening of atomic or molecular transitions can be extracted with SR spectroscopy probing emitters at a typical distance equal to the reduced optical wavelength (~$\lambda/2\pi$). This effective sub-wavelength confinement of the probed atoms allows for extremely high atomic densities with the additional benefit of sub-Doppler resolution. The collisional broadening ($\gamma_c$) obtained after fitting the experimental data at various pressures is shown in Fig. 5. After linear regression analysis we obtain a slope of 28(9) MHz/Torr (dashed black line). Ideally, the intercept of the curve with the zero pressure line should be given by the natural linewidth of the $6^2S_{1/2}\rightarrow5^2D_{5/2}$ transition that is 124 kHz, dominated by the electric-dipole $5^2D_{5/2} \rightarrow 6^2P_{3/2}$ transition [31]. However, our experiments show that linewidth tends to a value of 15(10) MHz at zero pressure. Beside the limited precision of the measurement, the main reason behind this discrepancy is the FM modulation amplitude depth of 10MHz limiting the minimal width of the observed spectrum. Other possible sources of broadening could be the laser low-frequency noise, as a fraction of a second is required to scan though an atomic transition, or the presence of impurities inside the cell inducing broadening due to velocity changing collisions [33]. Additionally, the long averaging times could induce broadening and lineshape distortion (due to uncertainties in the frequency scale calibration) that also contribute to systematic errors in the determination of the broadening. We note also that the minimum observed linewidth in SR spectroscopy has consistently been larger compared to its theoretical limitations [16] also confirmed by a comparison with saturated absorption experiments in the volume of the cell [34]. Our measurements indicate that the collisional shift is negligible for the range of pressures explored in this experiment, as can be observed on Fig. 4, where the SR resonances coincide with the expected hyperfine frequency positions (vertical dashed-dotted black lines) at both low and high temperature.

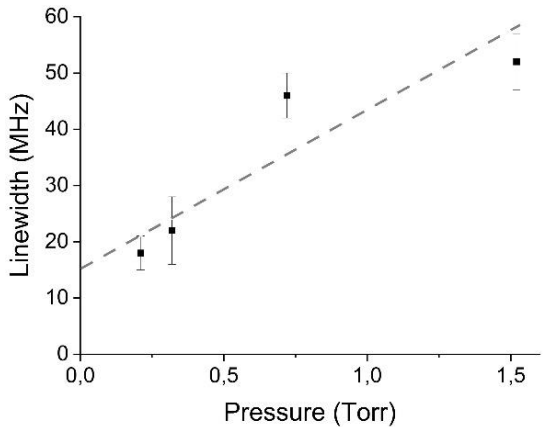

**Figure 5.** Width of the fitted Selective Reflection signal as a function of pressure. The linear fit, with a slope of 28(9) MHz/Torr and an intercept at ~15 MHz is also shown as a dashed black line.

The collisional broadening data are obtained by measurement of the real (dispersive) part of the effective vapor susceptibility [see eqs. (2) & (3)]. The results are in relatively good agreement with the absorption measurements reported in [35] providing a self-broadening of 37MHz/Torr on the $6S_{1/2} \rightarrow 5D_{5/2}$ line. However, in experiment of [35] the probed quantity is the vapor absorption in a macroscopic cell, namely the imaginary part of the vapor susceptibility, focusing on the analysis of the wings of the Doppler broadened spectra. Systematic uncertainties mostly related to the temperature measurement and homogeneity in our experiment could also account for the difference between the two reported values. We also note that in the range of cesium densities investigated in this experiment the effects of cesium dimers should be negligeable [36].

In order to fit the SR spectra of this experiments we have to account for the influence of atom-surface interactions. Indeed, Casimir-Polder measurements are a major motivation for SR experiments [25]. However, the signal to noise ratio is moderated in our experiment due to the very weak transition strength. In order to address this limitation, we used a large FM modulation amplitude and long integration times after tightly filtering the signal by applying large time constants on the lock-in detection. This causes inevitable distortion of the signal shapes and eventually increases the uncertainty of our $C_3$ coefficient measurement. Summarizing the extracted $C_3$ coefficient from our measurements at different densities we find that $C_3$=6(5) kHz.µm$^3$. This value is significantly different from the theoretical prediction of 0.15 kHz.µm$^3$ [37]. Apart from statistical and systematic errors due to the small signal amplitude, a possible explanation for this deviation is that the $5^2D_{5/2}$ atoms are sensitive to other parasitic interactions (patch potentials or charges trapped on the dielectric surface).

In conclusion, we present the first Doppler-free experiment involving a single electric-quadrupole transition. This has been achieved using the linear selective reflection technique. The main technical challenge was to extract an extremely small signal with an amplitude of ~30 parts per billion. The experiment provides information on the collisional broadening of the $6^2S_{1/2}$—$5^2D_{5/2}$ transition extracted from a dispersive, sub-Doppler measurement.

Our experiments pave the way for using further investigations of the Casimir-Polder interaction using quadrupole transitions. For example, improving the sensitivity of our selective reflection experiment could allow probing retardation effects on the excited state $5^2D_{5/2}$ as predicted in [37]. Furthermore, selective reflection on the electric-quadrupole $6^2S_{1/2}$—$6^2D_{3/2}$ line could allow investigations of the resonant atom-polariton coupling without an intermediate pumping step [18], [20] that further complicates the interpretation of spectroscopic Casimir-Polder measurements.

**Funding.** This work was supported by the Singapore Ministry of Education, grant No: MOE-T2EP50120-0005.

**Acknowledgements**
We thank the group SAI of the Laboratoire de Physique des Lasers for providing the cesium cell for these experiments.

**References**
1. V. S. Letokhov and Veniamin Pavlovich Chebotaev, Springer, Berlin, 1977 (n.d.).
2. W. Demtröder, Springer, Berlin, 2003 (Springer, Berlin, 2003, 2003).
3. A. D. Ludlow, M. M. Boyd, J. Ye, E. Peik, and P. O. Schmidt, Rev. Mod. Phys. **87**, 637 (2015).
4. V. Klimov, D. Bloch, M. Ducloy, and J. R. Rios Leite, Opt. Express **17**, 9718 (2009).
5. V. V. Klimov, D. Bloch, M. Ducloy, and J. R. Rios Leite, Phys. Rev. A **85**, 053834 (2012).
6. C. T. Schmiegelow, J. Schulz, H. Kaufmann, T. Ruster, U. G. Poschinger, and F. Schmidt-Kaler, Nat. Commun. **7**, 12998 (2016).
7. D. Antypas, A. Fabricant, J. E. Stalnaker, K. Tsigutkin, V. V. Flambaum, and D. Budker, Nat. Phys. **15**, 120 (2019).
8. C. S. Wood, S. C. Bennett, D. Cho, B. P. Masterson, J. L. Roberts, C. E. Tanner, and C. E. Wieman, Science **275**, 1759 (1997).
9. J. Guéna, D. Chauvat, Ph. Jacquier, E. Jahier, M. Lintz, S. Sanguinetti, A. Wasan, M. A. Bouchiat, A. V. Papoyan, and D. Sarkisyan, Phys. Rev. Lett. **90**, 143001 (2003).
10. K.-H. Weber and C. J. Sansonetti, Phys. Rev. A **35**, 4650 (1987).
11. F. Ponciano-Ojeda, S. Hernández-Gómez, O. López-Hernández, C. Mojica-Casique, R. Colín-Rodríguez, F. Ramírez-Martínez, J. Flores-Mijangos, D. Sahagún, R. Jáuregui, and J. Jiménez-Mier, Phys. Rev. A **92**, 042511 (2015).
12. E. A. Chan, S. A. Aljunid, N. I. Zheludev, D. Wilkowski, and M. Ducloy, Opt. Lett. **41**, 2005 (2016).
13. J.-L. Cojan, **9**, 385 (1954).
14. J. P. Woerdman and M. F. H. Schuurmans, Opt. Commun. **14**, 248 (1975).
15. A. M. Akul'shin, V. L. Velichanskii, A. S. Zibrov, V. V. Nikitin, V. V. Sautenkov, E. K. Yurkin, and N. V. Senkov, JETP Lett. **36**, 303 (1985).

16. M. Oria, M. Chevrollier, D. Bloch, M. Fichet, and M. Ducloy, Europhys. Lett. EPL **14**, 527 (1991).
17. M. Chevrollier, D. Bloch, G. Rahmat, and M. Ducloy, Opt. Lett. **16**, 1879 (1991).
18. H. Failache, S. Saltiel, M. Fichet, D. Bloch, and M. Ducloy, Phys. Rev. Lett. **83**, 5467 (1999).
19. H. Failache, S. Saltiel, A. Fischer, D. Bloch, and M. Ducloy, Phys. Rev. Lett. **88**, 243603 (2002).
20. A. Laliotis, T. P. de Silans, I. Maurin, M. Ducloy, and D. Bloch, Nat. Commun. **5**, (2014).
21. E. A. Chan, S. A. Aljunid, G. Adamo, A. Laliotis, M. Ducloy, and D. Wilkowski, Sci. Adv. **4**, eaao4223 (2018).
22. N. Papageorgiou, M. Fichet, V. V. Sautenkov, D. Bloch, and M. Ducloy, Laser Physics **4**, 392 (1994).
23. V. Vuletić, V. A. Sautenkov, C. Zimmermann, and T. W. Hänsch, Opt. Commun. **99**, 185 (1993).
24. D. Bloch and M. Ducloy, in *Advances In Atomic, Molecular, and Optical Physics* (Elsevier, 2005), **50**, pp. 91–154.
25. A. Laliotis, B.-S. Lu, M. Ducloy, and D. Wilkowski, AVS Quantum Sci. **3**, 043501 (2021).
26. J. Lukusa Mudiayi, I. Maurin, T. Mashimo, J. C. de Aquino Carvalho, D. Bloch, S. K. Tokunaga, B. Darquié, and A. Laliotis, Phys. Rev. Lett. **127**, 043201 (2021).
27. S. Tojo, M. Hasuo, and T. Fujimoto, Phys. Rev. Lett. **92**, 053001 (2004).
28. S. Tojo, T. Fujimoto, and M. Hasuo, Phys. Rev. A **71**, 012507 (2005).
29. E. A. Chan, S. A. Aljunid, G. Adamo, N. I. Zheludev, M. Ducloy, and D. Wilkowski, Phys. Rev. A **99**, 063801 (2019).
30. B. K. Sahoo, Phys. Rev. A **93**, 022503 (2016).
31. S. Pucher, P. Schneeweiss, A. Rauschenbeutel, and A. Dareau, Phys. Rev. A **101**, 042510 (2020).
32. M. Ducloy and M. Fichet, J. Phys. II **1**, 1429 (1991).
33. A. Laliotis, I. Maurin, M. Fichet, D. Bloch, M. Ducloy, N. Balasanyan, A. Sarkisyan, and D. Sarkisyan, Appl. Phys. B **90**, 415 (2008).
34. D. Bloch, P. Todorov, J. C. De Aquino Carvalho, I. Maurin, and A. Laliotis, in *20th International Conference and School on Quantum Electronics: Laser Physics and Applications*, T. N. Dreischuh and L. A. Avramov, eds. (SPIE, 2019), p. 60.
35. A. Andalkar, M. Iinuma, J. W. Smit, and E. N. Fortson, Phys. Rev. A **70**, 052703 (2004).
36. K. Niemax, J. Quant. Spectrosc. Radiat. Transf. **17**, 125 (1977).
37. J. C. de A. Carvalho, P. Pedri, M. Ducloy, and A. Laliotis, Phys. Rev. A **97**, (2018).